\documentclass{epl}

\title{Modelling background charge rearrangements near 
        single--electron transistors as a Poisson process}
\author{Heinz--Olaf M\"{u}ller\inst{1,2}\thanks{hom@ise.ch}, 
        Miha Furlan\inst{1,3}, 
        Thomas Heinzel\inst{1} \and 
        Klaus \mbox{Ensslin\inst{1}}}
\institute{
  \inst{1} Solid--State Physics Laboratory, ETH Z\"{u}rich -- 
                CH--8093 Z\"{u}rich, Switzerland\\
  \inst{2} Integrated Systems Engineering, ISE -- CH--8008 
                Z\"{u}rich, Switzerland\\
  \inst{3} Swiss Federal Office of Metrology and Accreditation, 
                metas -- CH--3003 Bern--Wabern, Switzerland
}

\shortauthor{H.-O. M\"{u}ller {\it et al.}}
\shorttitle{Modelling background charge rearrangements}

\begin{document}

\maketitle

\begin{abstract}
  Background charge rearrangements in metallic single--electron
  transistors are modelled in two--level tunnelling systems
  as a Poisson process with a scale parameter as only
  variable. The model explains the recent observation of asymmetric
  Coulomb blockade peak spacing distributions in metallic
  single--electron transistors. From the scale parameter we estimate
  the average size of the tunnelling systems, their density of states,
  and the height of their energy barrier. We conclude that the
  observed background charge rearrangements predominantly take place
  in the substrate of the single--electron transistor.
\end{abstract}

\section{Introduction}

% Intro: role of the SET
The metallic single--electron transistor (SET)~\cite{ful2,gra2} is a
possible building block of future electronics, based upon the
controlled transfer of individual charges onto and off small isolated
electrodes, called islands. However, due to the extreme sensitivity of
SETs to charges close to these islands, a static background charge
arrangement is essential for proper device operation~\cite{kor1,cov1}.
On the other hand, this very high sensitivity makes the SET an excellent
device to investigate background charge rearrangements, {\it e.g.}
in nearby two--level tunnelling systems (TLTS)~\cite{zim1,zor1}.
One issue of this paper is to determine the location of TLTSs within
the device.

% Our work
In recent experiments~\cite{fur1,fur2} on Al/AlO$_\textrm{x}$/Al SETs
it was shown that the distribution of nearest--neighbour spacings
(NNS) between Coulomb blockade oscillation peaks is influenced by such
charge rearrangements: they generate a pronounced tail in the NNS
distribution towards smaller spacings (see left inset of fig.~\ref{fig1}).
Here, we propose a quantitative model for this tail. We argue that
background charge rearrangements can be interpreted in terms of a
Poisson process. This way, we can fit the shape of the NNS
distribution tail, with a single scale parameter. From this fit, we
find an average hopping distance in the TLTS of about 4\,nm and
conclude that the charge rearrangements do not occur in the tunnel
barriers of the SET. We can furthermore estimate the density of states
of TLTSs and the barrier height between the two TLTS states.

\section{Experimental observations and statistical model}

% p(Vg) distribution
The NNS distribution $p(\Delta V_g)$ in fig.~\ref{fig1} is strongly
peaked at $\Delta V_g = e/C_g$, but displays a remarkable tail towards
lower $\Delta V_g$ as well, which contains up to one third of the
total data points. Here, $V_g$ and $C_g$ denote the gate voltage and
the gate capacitance, respectively. The spacings observed in the
low--$\Delta V_g$ tail are stable: repeated and reversed gate voltage
sweeps reveal the same statistical distribution of events. 
It has been attributed to the rearrangement
of background charges in TLTSs~\cite{lan1}, fig.~\ref{fig2}{\it a}).  
A further important experimental result is that the 
NNS distribution does not
depend on $V_g$: it was found~\cite{fur2} that the standard deviation
of the $n$--th neighbour Coulomb peak spacings is proportional to
$\sqrt{n}$.

% TLTS and random telegraph noise (RTN)
The clearly observed hysteresis of the Coulomb peak pattern when
reversing the gate scan direction~\cite{fur2} suggests the
existence of metastable configurations, which explains the high 
stability and reproducibility of the experimental results.  
Neither the Coulomb blockade peak positions nor the
underlying smooth $1/f$ noise spectrum depend on $V_g$ or the scan
speed.  Hence, other excitations which could spontaneously flip a TLTS
(like thermal activation) are much weaker compared to the parameters
which we control in the experiment. Note that these experiments are
different from dynamic and frequency dependent background charge noise
measurements~\cite{zim1,zor1,tlf1,tlf2,tlf3}.

% TLTS mechanism
The focus of our model lies on the $p(\Delta V_g)$ distribution
generated by TLTS fluctuations.  Since the $\Delta V_g$ distribution is
stable, a single sweep of $V_g$ already reveals it completely.
Therefore, our study is restricted to a single gate sweep.  During
this sweep, the electric field around the island electrode retains its
direction thus causing all background charges in TLTSs to `hop' either
towards or away from the island electrode. Therefore, the total amount
of charge $Q_i$ influenced on the island is monotonic as a function of
$V_g$ and is tested periodically (with respect to $V_g$) by the
position of the Coulomb peak.  Background charges in TLTSs contribute
to the variation of $\Delta V_g$ by influencing additional charge on
the central island.

% Transformation
We display the situation of the experiment in fig.~\ref{fig2}
schematically. While sweeping $V_g$, TLTS switching events lead to
stochastic jumps of the island charge $Q_i$, see 
fig.~\ref{fig2}{\it b}). Due to the discrete
technique of measuring the Coulomb spacings, the $Q_i$ jumps are
recorded at almost regular $V_g$ intervals. From a mathematical point
of view, we can consider $V_g$ as a function of $Q_i$ just as well. In
this picture, see fig.~\ref{fig2}{\it c}), 
we observe steps of $V_g$ of (almost) uniform height $\Delta V_g$ 
which occur at arbitrary values of $Q_i$.

% Poisson process
The sequence of $Q_i$ values, at which $V_g$ jumps, forms a stochastic
process where the charge $Q_i$ takes over the role of the time in the
common notation of stochastic processes. Assuming statistic
independence of the jumps (see below), the process can be identified
as Poisson process~\cite{kam1}. The single requirement for a Poisson
process is the independence of the individual events.  The process is
described by a single scale parameter $1/\gamma$. The probability that
$n$ jumps occur before the charge $Q_i$ is accumulated on the island,
is given by~\cite{kam1}

\begin{equation}
p_n(Q_i) = \frac{(\gamma\, Q_i)^n}{n!}\,\exp(-\gamma\, Q_i).
\label{dist}
\end{equation}

The Poisson process is non--stationary, which in our context
corresponds to an average number of jumps $\langle n\rangle = \gamma
Q_i$, which depends on $Q_i$. $\gamma Q_i$ also corresponds to the
square standard deviation, {\it i.e.}  $\langle\langle
n\rangle\rangle = \langle n^2\rangle - \langle n\rangle^2 = \gamma
Q_i$.

% Description of the tail
In the framework of the Poisson process it is possible to describe the
tail of the $p(\Delta V_g)$ distribution function. For
that purpose we transform $p(\Delta V_g)$ into $p(\Delta Q_i)$, which
follows from

\begin{equation}
\label{vg}
\Delta Q_i = e-\Delta V_g C_g.
\end{equation}

$p(\Delta Q_i)$ is peaked at $\Delta Q_i=0$ and displays a tail into
the positive, see right inset in fig.~\ref{fig1}. 
The probability distribution of the charge difference
$\Delta Q_i$ between two jumps corresponds to the waiting time
distribution of the conventional Poisson process. Therefore

\begin{equation}
\label{gamma}
p(\Delta Q_i) = -\,\frac{{\rm d}p_0(\Delta Q_i)}{{\rm d}\Delta Q_i}
 = \gamma\,\exp(-\gamma\Delta Q_i)
\end{equation}

is the tail of the distribution function $p(\Delta Q_i)$ with
$p_0(\Delta Q_i)$ as defined in eq.~(\ref{dist}).  Finally, the mean
value and square standard deviation of $\Delta Q_i$ also follow from
the distribution function $p(\Delta Q_i)$:

\begin{equation}
\label{deltaqi}
\langle\Delta Q_i\rangle =
\sqrt{\langle\langle\Delta Q_i\rangle\rangle} = \frac{1}{\gamma}.
\end{equation}

\section{Discussion}
\begin{figure}
  \onefigure{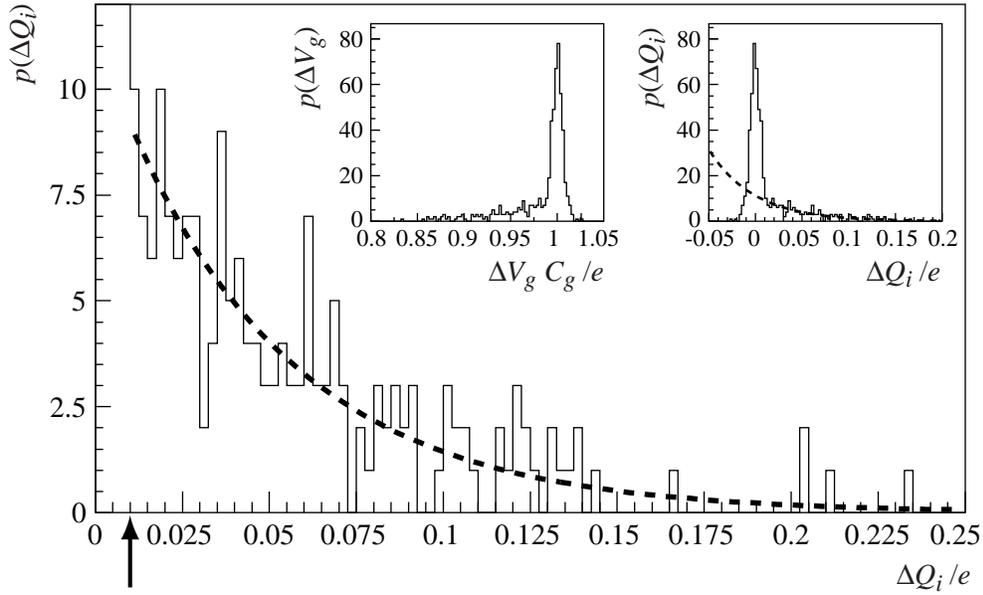}
  \caption{Determination of $1/\gamma$ from a fit to experimental
    data.  Equation~(\ref{gamma}) is used to fit the tail of the probability
    distribution function $p(\Delta Q_i)$.  The main plot is enlarged
    to emphasize the fit of the tail, whereas the left and right inset
    show the complete $p(\Delta V_g)$ and $p(\Delta Q_i)$ histograms,
    respectively. The dashed lines in the plots represent the fit, and
    the arrow indicates the limit of the fit range.}
  \label{fig1}
\end{figure}

\begin{figure}
  \onefigure{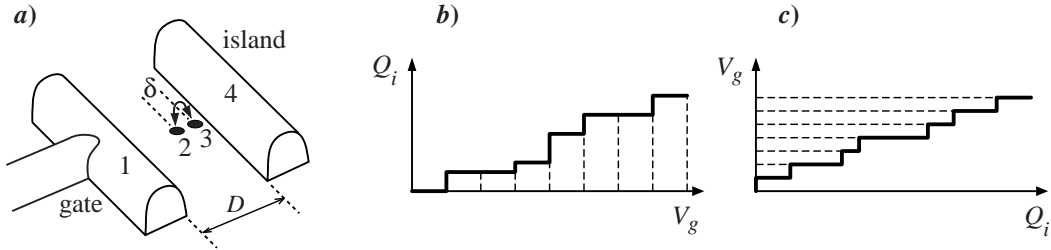}
  \caption{Background charges in two--level tunnelling systems. 
    {\it a})~Schematic drawing of the SET's gate (1) and 
    island (4) electrode, which are separated by the distance $D$.
    Because the island is
    quasi--isolated, the source--drain electrodes are not considered
    for electrostatic arguments. A TLTS is located between island and
    gate, where a charge can tunnel a distance $\delta$ from one site
    (2) to another (3) or vice versa. 
    {\it b})~The island charge $Q_i$
    displays arbitrary jumps at definite values of $V_g$. 
    {\it c})~Transition to uniform $V_g$ steps at 
    arbitrary $Q_i$ values.}
  \label{fig2}
\end{figure}

% Tail vs peak
It has to be noted that the presented stochastic model is only valid
for the description of the tail of the $p(\Delta Q_i)$ distribution,
but not its maximum.  While the tail is due to events arising from
randomly distributed TLTSs, for which the presumptions of the Poisson
process will be proven to hold, the maximum reflects the regularly
spaced Coulomb blockade peaks and is deterministic in nature.
Fortunately, these two ranges can easily be distinguished in our
experimental data (see fig.~\ref{fig1}).  In order to separate both
domains, a phenomenological cut is made at $\Delta Q_{i, {\rm min}} =
0.01\,e$, which corresponds via (\ref{vg}) to $\Delta V_{g, {\rm max}}
= 0.99\,e/C_g$.

% Values of gamma
Our model provides us with three independent methods of determining
the value of its sole parameter $1/\gamma$: (i) a fit to
eq.~(\ref{gamma}), (ii) the average $\langle\Delta Q_i\rangle$, and
(iii) the deviation $\langle\langle\Delta Q_i\rangle\rangle$ according
to eq.~(\ref{deltaqi}).  For the data presented in fig.~\ref{fig1} we
find the respective values: (i) $1/\gamma = (0.046 \pm 0.002)\,e$,
(ii) $1/\gamma = 0.057\,e$, and (iii) $1/\gamma = 0.043\,e$.  The
discrepancy between the average obtained from (ii) and the other
values can be shown to be due to the $\Delta Q_{i, {\rm min}}$ cutoff.
However, the correspondence between (i) and (iii) is sufficient to
demonstrate that the TLTS events are statistically independent because
it reflects a unique feature of the Poisson process. This process
provides an adequate description of $p(\Delta Q_i)$ and 
in turn $p(\Delta V_g)$.  Measurements on five different SET 
devices of equal geometry and on the same substrate result in a mean 
$1/\bar{\gamma} = (0.047 \pm 0.003)\,e$.

% The value of gamma
The value of $1/\gamma$ provides a charge scale of the influence of
TLTSs on the island charge $Q_i$. Thus, it describes the material
properties and consequences of the sample manufacture.  A small value
of $1/\gamma$ corresponds to a short tail of the distribution
functions $p(\Delta Q_i)$ and $p(\Delta V_g)$, which in turn
correlates with few TLTS transitions per $e/C_{g}$ interval and, thus,
rather stable device operation. $1/\gamma$ also scales with device
dimensions, as will be shown below.

% The link to tunnelling distance, assumptions
A simple geometric argument allows us to deduce a mean tunnelling
distance $\delta$ of the TLTS from $1/\gamma$.  It is based on the
``constant capacitance model''~\cite{kou2}: charge rearrangements
within TLTSs induce potential shifts only, but they do not alter any
capacitances.  This is appropriate here, since only metal electrodes
with high electron density and individual charges at the TLTS sites
are considered.  The assumption of constant capacitances is also
supported by the experimental data, since the peak position of
$p(\Delta V_g)$ is independent of $V_g$ and the bias voltage (see
figs.~1{\it d}) and 3{\it b}) in ref.~\cite{fur1}, respectively).  
In addition,
screening of the single TLTS charges is neglected, which is reasonable
considering the low concentration of quasi--free electrons around
them, {\it i.e.} in silicon oxide or aluminum oxide.  The treatment
is simplified by assuming a constant field between gate and island,
and by assuming that only one electron is rearranged in a TLTS
switching event.

% Green's reciprocation theorem
In this case, Green's reciprocation theorem~\cite{smy1} can be applied
to the setup displayed in fig.~\ref{fig2}{\it a}),
\begin{equation}
\sum_{j=1}^nQ_jV_j^{\prime} = \sum_{j=1}^nQ_j^{\prime}V_j,
\label{greens}
\end{equation}
where $\{Q_j\}$ and $\{Q_j^{\prime}\}$ are two charge configurations
in a system of conductors and $\{V_j\}$ and $\{V_j^{\prime}\}$ are the
respective voltage configurations.

% Start the derivation
$\{Q_j\}$ shall be the charge configuration before and
$\{Q_j^{\prime}\}$ that after charge transfer.  The influence on the
gate electrode is negligible, {\it i.e.} $Q_1 = Q_1^{\prime}$ and
$V_1 = V_1^{\prime} = V_{g}$.  Furthermore, we use a single charged
TLTS,

$$
Q_2 = e \hspace{1cm}
Q_2^{\prime}= 0 \hspace{1cm}
Q_3 = 0 \hspace{1cm}
Q_3^{\prime}= e.
$$

In our model the charge on the island is conserved, $Q_4=Q_4^{\prime}
= Q_i$. The island potential, however, shows a slight increase,
$V_4^{\prime} = V_4 + \Delta V_4$.  This yields with eq.~(\ref{greens})

\begin{equation}
\Delta V_4 = \frac{e}{Q_i}\,(V_3-V_2^{\prime}).
\label{volts}
\end{equation}

% Geometry dependence
For symmetry reasons the mutual contributions to the potentials $V_3$
and $V_2^{\prime}$ cancel and $(V_3-V_2^{\prime})$ is solely
determined by $V_g$,

\begin{equation}
\label{v3v2}
V_3-V_2^{\prime} = \frac{\delta}{D}\,V_g
\end{equation}

using the gate--island separation $D$ and the TLTS size $\delta$ (see
fig.~\ref{fig2}{\it a})). In general, eq.~(\ref{v3v2}) looks different for a
different electrode geometry, but a characteristic length scale
corresponding to $D$ can be found and a similar formula will hold in
most cases. The equation gives an upper estimate of the voltage
difference since it assumes the TLTS to live in the high--field
region. This region experiences the widest energy range of TLTS
transitions and thus contributes most events to the statistics.

% Assembling the terms
Finally, $Q_i = C_gV_g$ together with (\ref{volts}) and (\ref{v3v2})
yields

\begin{equation}
\label{delta}
\Delta V_4  = \frac{\delta}{D}\,\frac{e}{C_g}.
\end{equation}

This value is independent of $V_g$, as found in the experiment, due to
the cancellation of the linear $V_g$ dependence of $Q_i$ and
$(V_3-V_2^{\prime})$.

% The gamma delta connection
Fixed capacitances allow us to relate $\Delta V_4$ of eq.~(\ref{delta})
to $\Delta V_g$ of eq.~(\ref{vg}) and to establish a link between
$1/\gamma$ of (\ref{gamma}) and $\delta$ of (\ref{delta}). Given the
regular Coulomb peak separation $e/C_g$, the link is provided by

\begin{equation}
\label{deltagamma}
\frac{\Delta V_4}{e/C_g} = \frac{1/\gamma}{e}\hspace*{1cm}
\delta = \frac{D}{e\,\gamma}.
\end{equation}

% Missing distance
Hence, $\gamma$ scales linearly with $D$, {\it i.e.}
the NNS tail is experimentally observable only for sufficiently 
small $D$. However, the result
(\ref{deltagamma}) is independent of the distance between the TLTS and
the island.  This is due to the assumption that the charge of the TLTS
influences an equal amount of charge on the island, no matter what the
capacitance (and thus the distance) between them is. In other words,
this result follows directly from using the ``constant capacitance
model'' and neglecting screening. Although ref.~\cite{fur2} discusses
the TLTS influence in terms of a general model where the potential
rather than the charge of a TLTS is kept fixed, producing both a
distance and a size dependence of the TLTSs, it is not in
contradiction to the present treatment.  We rather consider the range
of small charge displacement and small distance from the SET island
here (top left part of fig.~5{\it b}) in ref.~\cite{fur2}). These
conditions are reasonable and have already been concluded in
\cite{fur1,fur2}. Both methods yield corresponding results in case of
short TLTS--island distance.

% Calculating delta
The values obtained for $1/\gamma$ easily translate into a
characteristic tunnelling distance $\delta$ within TLTSs, using
eq.~(\ref{deltagamma}).  With $D = 70 \ldots 100$\,nm, depending on the
device, we obtain $\delta = (3.94\,\pm\,0.19)$\,nm, taking into
account eight experiments with geometrically \emph{different} 
SETs measured at 10\,mK. This value agrees well with other 
experiments~\cite{zor1,fur1,fur2,kir1}.

% Shallow traps
The value we extract for $\delta$ thus exceeds the thickness of the
SET tunnel barriers, which is typically 2\,nm or less. Hence the
monitored trap states cannot reside in a tunnel junction itself, but
must be located in the surface oxide or the substrate~\cite{zor1}. The
measurements~\cite{fur1,fur2} suggest the tunnelling processes
themselves not to be field--assisted, since the probability of a TLTS
event does not depend on $V_g$.  Under these circumstances $\delta
\approx 4$\,nm is a rather large value.  In the case of tunnelling
between trap states, however, the larger tunnelling distance can be
compensated by lower barrier height, which can in fact be well below
1\,eV and still warrant electric field independence of the tunnelling
process.  Thus, the height of the energy barrier between the two TLTS
sites is limited towards low energies by the maximum potential
difference ($\approx80$\,meV) and towards high energies by the
necessary barrier transparency \mbox{($\approx0.7$\,eV)}. The neglect of
multiple electron transitions within one TLTS in our model might be
another reason for the large value of $\delta$.

% Temperature dependence
Information on TLTS switching events at elevated temperatures is hard
to access due to the increasing overlap between the main peak of the
$p(\Delta Q_i)$ distribution and its tail, as the temperature
increases.  Measurements performed at temperatures up to 200\,mK did
not reveal any significant variation of $1/\gamma$.

% TLTS density of states
The determination of $1/\gamma$ (and $\delta$) also provides an
estimate of the density of states of the TLTSs.  The data presented in
fig.~\ref{fig1} contain 572 Coulomb oscillation periods, 177 of which
belong to the tail of $p(\Delta V_g)$.  While the total energy range
scanned by $V_g$ is $572\,e/C_g\approx1.8$\,V, the potential shift
experienced between the two states of a TLTS is only $\delta/D$ times
this value. Hence, the average density of states is

$$
D_{\rm TLTS} = \frac{177}{572}\,\frac{D}{\delta}\,\frac{C_g}{e^2}
  \approx2.1\,\frac{1}{\rm meV}.
$$

Owing to lacking information on the energy level of the whole TLTS and
its inner structure (which might allow, for instance, more than one
TLTS transition), $D_{\rm TLTS}$ does not easily translate into a
density of states of the traps accommodating the TLTS.

% On dynamic fluctuations
Defects with internal degrees of freedom can switch between metastable
configurations also due to their interaction with a thermal bath or by
photon excitations.  Low--frequency excess noise even in high--quality
devices is significantly generated by dynamic fluctuations of such
defects~\cite{kir1}.  While a single fluctuator, also known as random
telegraph fluctuator (RTF), shows a Lorentzian spectrum, an ensemble
of RTF results in the typical $1/f$ noise~\cite{dut1}.  It is
reasonable and tempting to assume that the TLTS effects investigated
in our studies are of the same origin as RTF generated noise.  Due to
the metastable nature of the TLTS, the dynamics of a fluctuator
crucially depend on its energy threshold separation.  However, this
is only vaguely known from our experiments via the observation of
Coulomb peak pattern hysteresis~\cite{fur2}.  
Furthermore, it is not clear how to extrapolate the data, 
where we can resolve TLTS events with a large hysteresis,  
towards a small energy interval of $k_{B}T$, {\it i.e} for 
fluctuators of small time constants contributing to non-zero 
frequency noise.  
Therefore, a decisive conclusion on consequences of our
results for dynamic fluctuations and noise is too speculative at the
moment.  However, his subject is under further investigation.

\section{Conclusion}
We have investigated charge transfer in two--level tunnelling systems
(TLTSs).  The study is based on measurements performed on metallic
SETs.  The charge transfer can be described in terms of a Poisson
process, where the influenced charge $Q_i$ replaces the usual time
variable.  Thus, the prominent tail of the probability distribution
function $p(\Delta V_g)$ is governed by a single scale parameter
$1/\gamma$.  We can associate this parameter with a length scale
$\delta$ of the TLTS. We find $\delta = (3.94\,\pm\,0.19)$\,nm for the
samples of different geometries under consideration.  From this result
we conclude that the SET tunnel barriers are in our case free of TLTSs. 
The fluctuators  reside in the oxide around the island electrode
or in the substrate.  The height of the energy barrier separating two
TLTS states is between $80$\,meV and 0.7\,eV and the density of states
of TLTS is found to be $2.1$\,(meV)$^{-1}$.

\acknowledgments
We are indebted to Sergey~V.~Lotkhov for the excellent sample 
manufacture, as well as to Mathias Wagner and Ulrich Z\"{u}licke 
for thorough reading of the manuscript.

\end{document}